\documentstyle[aps,prl,psfig,multicol]{revtex}
\title{Self-organization   
in BML Traffic Flow Model: Analytical Approaches
\footnote{This  work was done in 1995, accepted for publication in 1996.
}}
\author{Y. Shi
\footnote{Present address: TCM, Cavendish Laboratory, Cambridge CB3 0HE, UK.
Email: ys219@cam.ac.uk. }}
\address{Department of Physics, Fudan University, Shanghai 200433, China}
\date{Published in {\em Commun.  
Theor. Phys.}  {\bf 31}, 85-90 (1999)}
\begin{document}
\draft
\maketitle
\begin{abstract}
Analytical investigations are made on BML 
two-dimensional traffic flow model with alternative movement and
exclude-volume effect.
Several exact results are obtained, including the 
upper critical density above which there are only jamming configurations
asymptotically,
and the lower critical density below which there are only moving 
configurations asymptotically. The jamming transition observed in the ensemble
average velocity  takes place at another 
critical density $p_{c}(N)$, which is dependent on the lattice size $N$ and
is in the intermediate region between the lower and upper critical densities. 
It is suggested that $p_{c}(N)$ is proportional to a power of $N$, 
in good agreement with the numerical simulation. 
The order parameter of this jamming transition is identified.
\end{abstract}
\begin{multicols}{2}
\section{Introduction}

        In recent years much attention is  paid on cellular 
automaton  models for the  investigations on  complex systems.  
These models can be viewed as statistical models with 
dynamics. Some models of traffic flow are under intensive studies.
A two-dimensional  model was introduced by Biham, Middleton
and Levine (BML) \cite{biham}. It 
is defined  on a square lattice with periodic boundary condition. 
Each site contains either
an arrow pointing upwards or  to the right, or is empty. 
The dynamics is controlled by the traffic light, such that the right arrows
move only on even time steps and the up arrows move on odd time steps. On 
even
time steps, each right arrows moves one lattice constant to the right unless
the site on its righthand side is occupied by an arrow, which is
either up or right. If
it is blocked, it does not move, even if during the
same time step the blocking arrow moves out of that site.
Similar rules apply to the up arrows, which move upwards. 
The velocity $v$ of {\em a} right (up) 
arrow is defined to be the number of  moves it makes within  a 
number of even (odd) time steps  divided by this 
number of time steps.
It has a maximal value $v\,=\,1$, indicating that the arrow
is never stopped. The minimal value 
$v\,=\,0$ represents that  the arrow is stopped during  the 
entire time duration. The average velocity $\overline{v}$ 
for the system is obtained by 
averaging $v$ over {\em all} the arrows in the system.
If it is further averaged over many asymptotic configurations, one
obtains the ensemble average. 

         BML model is fully deterministic. It is called  to be 
self-organized because whatever the initial condition    
is, one {\em often} (but not always)  finds 
in the {\em asymptotic} configurations that,
 {\em all} the arrows move freely in 
their turns
hence the velocity averaged over all the arrows is $\overline{v}\,=\,1$, 
or they  
are {\em all} stopped, with $\overline{v}\,=\,0$.
 These two types of configurations are referred
to as moving and jamming ones, respectively. In the language of dynamics, 
they are  the  biggest basins of 
attraction. 
Which   asymptotic configuration  is finally  reached depends on 
both the density of arrows and the initial condition.
Is there any asymptotic 
configuration in which some arrows are moving while others are 
blocked? The answer is yes. Consider a column 
occupied by  less than $N/2$ up arrows, where $N$ is the number of the 
lattice points on each column,   while  its 
two neighboring columns  are both  full of up arrows.
Then asymptotically the arrows in this column are moving forever, 
independent of other arrows, which are  all blocked. 
But such configurations are 
 rare, that is, they occupy a very small  volume 
 in the phase space,  compared 
with those  of the  moving or
jamming configurations.
This is indicated by the  simulation result  that 
there is a sharp  moving-to-jamming transition with the 
increase of arrow density. 
The simulation result (see Fig.3 of Ref.\cite{biham}) also 
tells  us that the fraction of phase space volume occupied by
the  moving or jamming configurations 
increases  with the size of the lattice. 

Here we make some analytical approaches. 
We give exact results on the lower  
critical density, below which there are only moving configurations
asymptotically,
 and an upper critical density, above which 
there are only jamming configurations asymptotically.
 Between these two critical
densities, the asymptotic configuration can be moving or jamming, or
even with both moving and blocked  arrows, depending
on the initial configurations. 
As indicated in the simulation, there is another
critical density above which the asymptotic configurations are 
 typically (but not always) jamming. 
This is the sharp (but not absolutely stepwise) jamming
transition discovered in the ensemble average velocity.
 
The content of this article is as follows. 
For convenience of discussions,  we
introduce some notations in Sec. II. In 
Secs. III and IV, we give some exact results,  
 the upper and lower critical
densities are determined. In Sec. V, by 
considering the typical 
pattern formation of the jamming cluster, we obtain, in a heuristic way,
 the    critical density for the  jamming transition.
The dependence on
the lattice size is determined, and
the order parameter is identified. Sec. VI is a summary.

\section{Notations}

For convenience of discussions, we introduce 
some notations here.
There are  $N\times N$ lattice points, the density of up (right)
arrows is $p_{\uparrow}\,=\,n_{\uparrow}/N^{2}$ 
($p_{\rightarrow}\,=\,n_{\rightarrow}/N^{2}$), where $n_{\uparrow}$ 
($n_{\rightarrow}$) is the number of up (right) arrows. The total density  
of arrows
is $p\,=\,p_{\uparrow}\,+\,p_{\rightarrow}$. The number of empty lattice
points is denoted as $n_{0}$.
The empty sites can be regarded as occupied by holes.
Each lattice point is given 
a coordinate $(i,j)$. $i$ and $j$ each runs from $1$ to $N$, hence
the lower-left corner is $(1,1)$. The periodic boundary
condition can be expressed as
\begin{equation}
(i+N,j)\,=\,(i,j+N)\,=\,(i+N,j+N)\,=\,(i,j).
\end{equation}
Because of the periodic boundary condition, 
 the lattice can be transformed 
in the way shown in Fig. 1, 
so that it  can  be 
viewed as a  parallelogram,  also with the periodic boundary condition.
This parallelogram   is made up of  $N$ lines parallel to the left-falling
diagonal
of the original square, on each of these lines
there are also 
$N$ lattice points. 
For convenience, we say that
 the lattice is composed of
$N$ ``left-falling diagonals'' (LFD). For example, 
this transformation translates the   line linking  
$(1,i)$ and $(i,1)$ $N$ units upwards 
to be connected with  the line linking $(i+1,N)$ and $(N,i+1)$,
composing   a  LFD.  Since the arrows
are right or up, this viewpoint is very useful in our discussions.

To avoid confusion, the word ``state'' is used for the lattice points, while
``configuration'' is for the whole system. The state of $(i,j)$ is
denoted as $|i,j\rangle$. $|i,j\rangle\,=\,\uparrow,\,\rightarrow$ or $0$ if 
 $(i,j)$ is occupied by an up arrow, a right arrow or a hole, respectively. 
 $|i,j\rangle$ is, of course, dependent on time, so it can be written as
 $|i,j\rangle(t)$ if necessary.
Obviously, in a moving configuration, 
 $|i,j\rangle(t)\,=\,|i+\delta,j\rangle(t+\delta)$ if $|i,j\rangle(t)\,=\,\rightarrow$,
 $|i,j\rangle(t)\,=\,|i,j+\delta\rangle(t+\delta)$ if $|i,j\rangle(t)\,=\,\uparrow$. 

\begin{figure}
\begin{picture}(200,400)
\setlength{\unitlength}{1pt}
\put(60,10){\line(1,0){150}}
\put(60,10){\line(0,1){150}}
\put(60,160){\line(1,0){150}}
\put(210,10){\line(0,1){150}}
\put(60,160){\line(0,1){150}}
\put(60,310){\line(1,-1){150}}
\put(60,130){\line(1,-1){120}}
\put(60,190){\line(1,0){120}}
\put(60,160){\line(1,-1){150}}
\put(40,304){$S'$}
\put(40,184){$O'$}
\put(40,154){$P$}
\put(40,124){$S$}
\put(40,4){$O$}
\put(220,4){$R$}
\put(220,154){$Q$}
\put(180,-5){$T$}
\put(198,184){$T'$}
\end{picture}
\vspace{0.2cm}
\caption{Equivalent transformation of the appearance of the lattice.
The square $PORQ$ can be transformed to the parallelogram $PRQS'$, because of
periodic boundary condition. The 
coordinates of the marked points are  $O(1,1)$, $R(N,1)$, 
$Q(N,N)$, $P(1,N)$, $S(1,N-1)$, $T(N-1,1)$, $O'(1,N+1)$,
$T'(N-1,N+1)$, $S'(1,2N-1)$. 
$O$ and $O'$, $T$ and $T'$, $S$ and $S'$ are
equivalently  the same points, respectively.}
\end{figure}
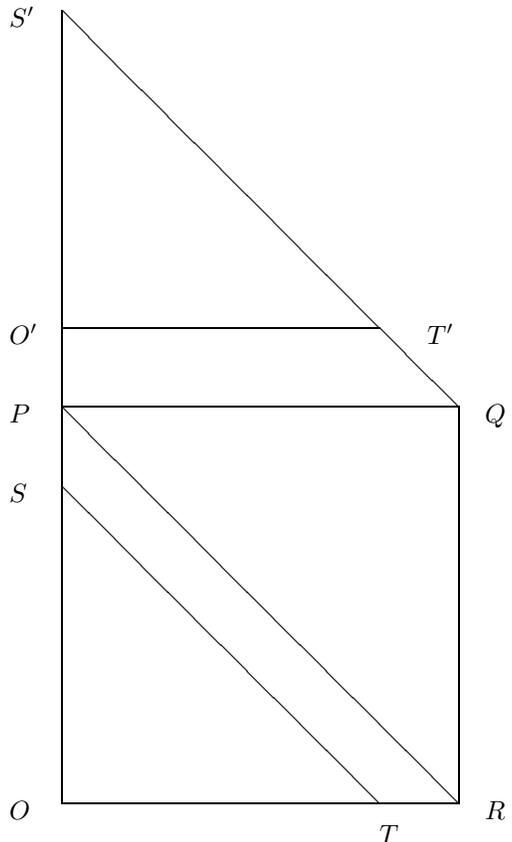

\section{Exact results on moving configuration}

First we point out that not only a jamming configuration, but also a moving
configuration is stationary, in the sense that 
all arrows of the same type  move simultaneously and 
thus form a rigid body.

The exact results are stated in the form of  theorems.

{\bf Theorem  1.}---{\em In a moving configuration, 
  a   LFD always consists of a same type of arrows, as well as holes.} 
                                                   
{\em Proof.}---Suppose $|i,j\rangle(t)\,=\,\uparrow$, while  
$|i-\delta, j+\delta\rangle(t)\,=\,\rightarrow$, 
where $\delta$ is a positive integer. 
If $t$ is an odd (even) time step, then 
after $\delta$ odd (even) time steps, the  right (up) arrow is blocked
by the up (right) arrow. This should not happen in a moving configuration. 
Because of periodic boundary condition,
every lattice point  on the same LFD as   $(i,j)$
can be represented as  $(i-\delta, j+\delta)$ with $\delta\,>\,0$. 
Therefore there
cannot be both up and right  arrows on a same LFD.
 Q.E.D.

{\bf Theorem  2}.---{\em In a moving configuration where  there are
 both right and up arrows,
there is at least one
empty LFD  at any instant.}

{\bf Proof.}---Without lose of generality, consider  an odd time step. 
For a LFD of up arrows, there cannot be any right arrow on its upside LFD,
seen as follows.
Suppose there are right arrows on this  upside  LFD. 
If a  right arrow is  just above an up arrow, the latter is blocked, 
which is not permitted in a moving configuration.
If a right arrow is above a hole on the   LFD composed of
 up arrows and
holes, at the next time step, this  right arrow will fill the 
hole (which has  moved one step upwards),  and  join the up arrows
(which have moved one step upwards)  on a same LFD. 
Hence there appears a LFD where there are both up and right arrows. 
This is forbidden in a moving configuration, according to Theorem 1.
If there are  both up and right arrows on the lattice, because of periodic
boundary condition, there must be right arrows ``before'' the up arrows, even 
though on the original square lattice they are ``after'' the up arrows. 
Therefore  there
is at least one empty LFD. On the other hand, 
at this  odd time step, for a LFD of right arrows, 
it is not necessary for its righthand side 
(just the upside) LFD to be empty, since the right arrows do not move at
this time step.
In conclusion,  the least number of empty LFD is only  $1$. Q.E.D.

{\bf Theorem 3}.---{\em Consider
 $N\,>\,2$.  There is an upper critical density, above which there is
no moving configuration.
The upper critical density is
 $1/2$
 if $N$ is 
odd, and is $1/2-1/2N$ if $N$ is even. }

{\em Proof.}---For $N\,=\,2$,
 there cannot be any moving configuration 
with the presence of both up and right arrows.
Hence we only consider  $N\,>\,2$. 
Without lose of generality, consider an odd time step.
At this time step, there 
can be an arrow on the righthand side of a right  arrow, but
 there cannot be any arrow on the upside of an  up arrow.
The most  crowded  configuration is the following:
an empty LFD is  on the upside of   a block of LFD-s consisting of up arrows
and holes, which is followed by a block of LFD-s consisting of right arrows
and holes.
Hence  the number
of up arrows $n_{\uparrow}$ and the holes in the up block, $n_0^{(\uparrow)}$,
 should satisfy $n_{\uparrow}\leq n_0^{(\uparrow)}$ if there are even
number of LFD-s in this block, and $n_{\uparrow}\leq n_0^{(\uparrow)}+N$ if 
there
are odd  number of LFD-s in this block. 
Similarly,   the number of right arrows   $n_{\rightarrow}$ 
and the holes in the right  block, $n_0^{(\rightarrow)}$, 
 should satisfy $n_{\rightarrow}\leq n_0^{(\rightarrow)}$ if there are even
number of LFD-s in this block, and $n_{\rightarrow}\leq n_0^{(\rightarrow)}+N$
 if there
are odd  number of LFD-s in this block.
In addition, the total  number of the holes on the lattice should 
satisfy 
$n_0\geq n_0^{(\uparrow)}+n_0^{(\rightarrow)}+N$, since there is at
least one empty LFD, according to Theorem 2. Therefore if $N$ is even, at the most crowded case,
we have odd number of non-empty LFD-s, hence we have either an 
odd number of up LFD-s and an even number of right LFD-s, or  an 
even number of up LFD-s and an odd number of right LFD-s.
In either case, we have 
$n= n_{\uparrow}+n_{\rightarrow}$$\leq n_0^{(\uparrow)}+n_0^{(\rightarrow)}+N$
$\leq n_0$$=N^2-n$, hence $p=n/N^2$$\leq 1/2$.
If $N$ is odd, at the most crowded case,
we have even number of non-empty LFD-s, hence we have either an 
odd number of up LFD-s and an odd number of right LFD-s, or  an 
even number of up LFD-s and an even number of right LFD-s.
In the odd-odd case, we have 
$n= n_{\uparrow}+n_{\rightarrow}$$\leq n_0^{(\uparrow)}+n_0^{(\rightarrow)}+2N$
$\leq n_0+N$$=N^2+N-n$, hence $p=n/N^2$$\leq 1/2+1/2N$.
In the even-even case, we have 
$n= n_{\uparrow}+n_{\rightarrow}$$\leq n_0^{(\uparrow)}+n_0^{(\rightarrow)}$
$\leq n_0-N$$=N^2-N-n$, hence $p= n/N^2$$\leq 1/2-1/2N$.
Combining these two cases we have $p\leq 1/2-1/2N$ if  $N$ is even.
Q.E.D.

\section{Exact results on jamming configuration}
Because  an up arrow can only be blocked by an  arrow (which can be 
right or up) above it,
while a right arrow can only be blocked by an arrow on its  righthand side,
 these arrows form a directed path in a jamming configuration.
All directed paths  point upwards or to the right.
Considering the 
periodic boundary coundition, one may obtain 
the following thorems.

{\bf Theorem  4}.---{\em 
In a jamming configuration, starting from an arbitrary arrow, 
one can obtain  a directed 
path which returns  to either the starting arrow or  
another  arrow on this path.}

We call such a path  a closed path. If it returns to the starting arrow, 
it is  a circular path.   Each
closed path contains  a circular path as  a part. 

{\bf Theorem 5.}---{\em In a  jamming configuration, there must be at least  
one circular path.} 

Clearly this
 is a  necessary condition  for a configuration to be jamming.   

{\bf Theorem 6.}---{\em 
The length of a circular path is  $N$ if it is composed  of
only one type  of arrows, and  is $2N$ if it is composed 
 of both types of arrows.
 Here the unit of the   length is 
the lattice constant. For example, the length of an edge of the square
is $N-1$.
}

{\bf Proof.}---Obviously 
the  circular path made up of one type  of arrows is 
parallel to the edge of the square, hence its length is $N$.
 If the circular path is made up of
both types of arrows, because  it is directed, generally
it  appears as two parts on the square lattice. For
example,  one part is a directed path connecting   
$(1,J)$ and $(I,N)$, another part 
is a directed path connecting $(I,1)$    and $(N,J)$ .
Note that $(N,J)$ is a nearest neighbor of $(1,J)\equiv (N+1,J)$, and
$(I,N)$ is a nearest neighbor of $(I,1)\equiv (I,N+1)$.
Clearly the total length of such a circular path is
$2N$. 
Q.E.D.

{\bf Theorem 7}.---{\em 
There is a lower critical density, below which there is 
no jamming configuration. 
The lower critical density is
$(1+p_{s}/p_{l})/N$, where 
$p_{s}$ and $p_{l}$ are respectively the smaller and larger one of  
$p_{\uparrow}$ and $p_{\rightarrow}$. }

{\em Proof.}-Suppose there 
is a circular path made up of only the arrows with the larger density,
and there are no other arrows of this type.
The arrows with the smaller density  are  blocked by this circular path. 
Therefore $N_{l}\,=\,N$, $N_{s}\,=\,(p_{s}/p_{l})N$. 
Thus  $(1+p_{s}/p_{l})/N$ is the smallest possible 
density for the jamming 
configuration if  the circular path is made up of only one type of arrows.
If all the up and right arrows  take part in composing the
 circular path, the density is
$2/N$. Since  $2/N \geq (1+p_{s}/p_{l})/N$, the theorem holds in
general.
Q.E.D.

\section{Formation of a jamming configuration}
The so-called jamming transition  observed  in 
the simulation 
referrs to a sharp, but not stepwise,
 transition in the ensemble average velocity. 
There are both 
moving and jamming configurations between the upper and lower critical 
densities, depending on the initial condition.
In \cite{biham}, the critical
 density for the jamming  transition was defined to be at the 
center of the transition region.
Here we define the critical density of jamming transition as 
the
value of the density at which the ensemble average
 velocity starts to be nearly
zero. 
We
approximate the non-stepwise 
transition in the ensemble average by a typical 
process of formation of a jamming configuration. Consequently
 the critical density of jamming transition is approximated 
 as the density above which there forms
a typical  jamming cluster.

A  jammming configuration is formed 
soon after the appearance of a  circular path, which  usually consists of 
both up and right arrows.  When an arrow meet the circular path, it is
  blocked. 
The circular path blocks the  right arrows on 
its lefthand side and  up arrows on its
downside. 
Blocked arrows block other arrows, and so on.
Consequently, a global cluster with directed branching structure emerges. 
The skeleton of this cluster is a  circular path. 

Now note that 
in the final jamming cluster, there are some 
arrows which are the ends of the cluster.
If the end-arrow is  an up (right) one, its upside (righthand side)
must be occupied, while 
there 
must be neither 
 right arrow on its lefthand side nor up  arrow on its 
downside.
 So the
 density of end-arrows are 
\begin{equation}
\rho_{e}(p)\,=\,p^{2}(1-p_{\rightarrow})(1-p_{\uparrow})\,
=\,p^{2}\,-\,p^{3}+p^{2}p_{\rightarrow}p_{\uparrow}\,\approx\,p^{2}.
\end{equation}
Hence the number of ends is $p^2N^2$.
Since the length of the circular path is $N$, it is very reasonable to 
assume 
that the number of the ends  is  a power of $N$ .
Consequently
the critical density for the ``jamming transition''
is a power of $N$, that is, 
\begin{equation}
p_{c}(N)\,=\,CN^{\alpha}, \label{alp}
\end{equation}
where $C$ is a coefficient, while $\alpha$ is the exponent.

The simulation results can be used to test this heuristic argument,  
and determine   
$\alpha$ and $C$. With the approximate values of $p_{c}(N)$ for $N\,
=\,16,\,32,\,64,\,128,\,512$ obtained from 
Fig. 3 in Ref. \cite{biham},
we obtain a good fit to Eq. (\ref{alp}) with $\alpha\approx -0.14$ and
$C \approx 0.76$, as shown in Fig.2.

\begin{figure}
\vspace{-5cm}
\psfig{figure=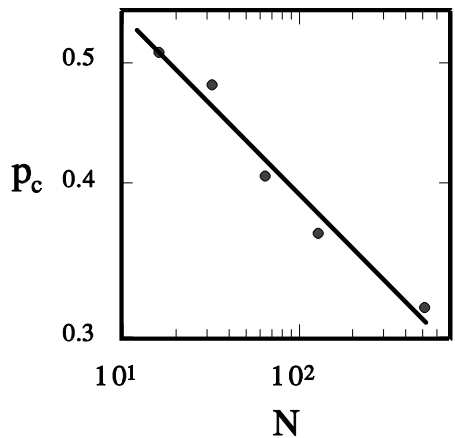}
\caption{Log-log plot of $p_{c}(N)$ with the lattice size $N$. $p_{c}(N)$ is
the  density 
above  which typical jamming configurations form, and thus the 
ensemble average velocity is nearly $0$.
The circles are the results  observed from Ref. [3]. The straight
line is the least square fit yielding $p_{c}(N)\,$$\approx\,CN^{\alpha}$,
with $C\approx 0.76$, $\alpha\approx -0.14$.}
\end{figure}
                           
Eq. (\ref{alp}) suggests that the jamming cluster at $p_{c}$ is a
fractal 
with dimensionality $2+\alpha\,=\,1.86$, which is close to 91/48, 
the fractal dimension of  
the infinite cluster of
two-dimensional
percolation \cite{feder}. This is understandable
 since the jamming cluster forms 
soon after the circular path forms, which is similar to the formation of
an infinite cluster in   percolation. The order 
parameter of percolation is the probability that an arbitrarily chosen
occupied site or bond belongs to an infinite cluster. Likewise,
we identify the 
the order parameter in jamming transition as
the probability that an arbitarily chosen
arrow belongs to a closed path. 

\section{summary}
 We have studied  BML two-dimensional traffic flow model analytically.
In particular we gives exact results on the two most typical asymptotic
configurations, the   moving configuration and jamming configuration.
In a moving configuration, all arrows keep moving, while in a jamming 
configuration, all arrows are blocked. 
Theorems 1 and 2 give two basic properties of a moving configuration. Based on
these, Theorem 3 provides the upper critical density, above which there is
no moving configuration asymptotically.
 The upper critical density is $1/2$ if $N$ is odd,
and is $1/2-1/2N$ is $N$ is even.  
Theorems 4, 5 and 6 give basic properties of a jamming configuration. The
crucial thing is the formation of a
 so-called circular path which cuts the lattice into two 
parts. Theorem 7 then gives the lower critical density, below which there
is no jamming configuration asymptotically. 
The so-called jamming transition observed in
the ensemble average velocity happens in a short  region between 
the upper and lower critical density. We define the critical density for the
jamming transition as that on which the emsemble average velocity begins to 
be close to zero. We investigate it approximately by considering the formation
of a typical jamming cluster. The $N$-dependent critical density $p_c(N)$
is claimed to be $CN^{\alpha}$, which fits  the numerical data very
well, with $C\approx 0.76$ and $\alpha\approx -0.14$.

The jamming transition turns out to be quite similar to percolation, and we
identify the order parameter as the probability that an arbitrarily chosen
arrow belongs to a closed path. Further investigations on  possible
criticality are interesting.

\end{multicols}


\begin{references}

\bibitem{biham}  O. Biham,  A. A.  Middleton  and  D. Levine, 
Phys. Rev. A
{\bf 46} (1992) 3290.

\bibitem{feder} D. Stauffer, {\em Introduction  to Percolation
Theory} (Taylor \& France, London, 1995). 

\end{references}
\end{document}